# Structural-disorder and its effect on the mechanical properties in single-phase TaNbHfZr high-entropy alloys

Soumyadipta Maiti [a, *], Walter Steurer [a]

[a] *Laboratory of Crystallography, Department of Materials, ETH Zurich, Vladimir-Prelog-Weg 5, CH-8093 Zurich, Switzerland*

## Abstract

Equiatomic TaNbHfZr refractory high-entropy alloys (HEAs) were synthesized by arc-melting. The HEAs were annealed at 1800°C for different times, at maximum up to 8 days. Their on average body-centered cubic (bcc), solid-solution structure was confirmed by X-ray (XRD) and neutron (ND) diffraction, respectively. The HEAs are characterized by high average values of the static atomic displacements from the ideal lattice positions and the local internal strain. The short-range clustering (SRC) of a subset of atoms, enriched in Hf and Zr takes place perpendicular to the <100> directions. Furthermore, it becomes increasingly interconnected as a function of the annealing time. This is revealed by the evaluation of diffuse XRD intensities, high-resolution transmission electron microscopy (HRTEM) images, and atom probe tomography (APT). The local structural disorder and distortions at the SRC were modeled by molecular dynamics (MD) relaxations. The hardness and compressive yield strength of the as-cast HEA is found to be many times of what can be expected from the rule of mixture. The yield strength further increases by 76 % after 1 day of annealing, which can be explained by a strengthening mechanism resulting from the SRC. With further annealing to 4 days, a minor phase with hexagonal close packed (hcp) structure and rich in Hf and Zr nucleates at the larger connecting nodes of the SRCs.

[*] Corresponding author: S. Maiti, Tel: +41 44 633 71 29, E-mail: smaiti@mat.ethz.ch

# 1. Introduction

Ideal high-entropy alloys (HEAs) are single-phase alloys of multiple principal elements in equal/near-equal proportion [1] with high configurational entropy increasing as $R\ln(n)$, where $R$ is the universal gas constant, $n$ being the number of solute elements. At high enough temperatures, this high entropic contribution to the Gibbs free energy of a HEA suppresses the formation of intermetallics and stabilizes a simple on average body-centered or face-centered cubic (bcc and fcc) phase [2, 3]. The broad distribution of the atomic radii of the different constituting elements leads to lattice distortions [4], and lowers the diffusion rate by an order of magnitude [5]. This in term gives rise to many interesting properties like high strength at ambient and high temperatures [6], diffusion barrier properties [7], fracture resistance at cryogenic temperatures [8] etc.

Of late, the use of refractory-element based HEAs (Mo, Nb, Ta, W and V) was introduced by Senkov et al [9, 10] mainly as candidates for high-temperature (HT) structural materials above 1100°C, especially for aerospace applications. Subsequently, refractory HEAs of lower density were investigated (W and Mo replaced with Hf, Zr, Ti and Cr) and found to possess excellent HT and ambient-temperature mechanical properties [11-13]. Moreover, refractory metals were suggested for the use in many diverse ambient-temperature applications like fabrication materials for micro- and nanoelectromechanical devices, electronics components and medical implants [14-16].

For most of the refractory HEAs, investigations were limited to the structures of as-cast alloys or inadequately annealed materials. Their long-term phase stability at elevated temperatures and the related mechanical performance have not been investigated in detail with a few exceptions (MoNbTaW HEA) [16, 17]. However, ample research has been done on both the experimental and theoretical aspects of disorders like short-range ordering and clustering (SRO and SRC) as well as Guinier-Preston zones (GP zones) in high-concentration solid-solution alloys since the 1960s [18-21], mainly based on X-ray and neutron diffuse scattering experiments. As expected, the SRO (SRC) has been found to largely influence the mechanical properties and the long-term structural stability of many engineering materials like the light-weight high-strength Al alloys [22], bcc Fe-Cr [23] and fcc austenitic steels [24-26] and nickel based alloys [27, 28] etc.

Since refractory HEAs have been found to be potential high-performance materials for both high- and ambient-temperature applications, we investigated an equiatomic single-phase TaNbHfZr HEA for its local structural disorder and its evolution with thermal treatments and ambient-temperature mechanical property changes.

## 2. Experiments

To synthesize the TaNbHfZr HEAs, first high-purity Ta and Nb powders (99.98 % and 99.99 % pure) were compacted and arc-melted under Ar atmosphere. Then, solid pieces of Hf and Zr (99.8 % pure, oxygen level ~ 150 ppm) were pre-alloyed to form a droplet and finally all the four elements were arc-melted together to form a cast of 6 grams. A Ti getter was used to consume traces of oxygen for all melting steps and the TaNbHfZr cast was re-melted five times, each time flipped upside down. The cast alloys were homogenized and subsequently annealed at 1800°C for 6 and 12 hours, 1, 2, 4 and 8 days, respectively, inside sealed Ta ampoules. This annealing temperature is 0.78 times the expected melting temperature of the alloy system [9] as per the rule of mixture. During long annealing treatments, an extra Hf-Zr oxygen-getter, wrapped in Ta foil, was provided inside the sealed ampoules to consume any traces of oxygen. For the ease of indicating the different annealing treatments of the alloys, they are termed as HEA-0D, HEA-1D, HEA-2D and so on for the as-cast, 1 and 2 days annealed material, respectively. Similarly, the alloys annealed for 6 and 12 hours are termed as HEA-6H and HEA-12H, respectively.

To determine the average structure and the phase-purity of the different annealed samples, powder XRD patterns were measured by an in-house powder diffractometer (PANalytical X'Pert PRO diffraction system) using Cu $K_{\alpha 1}$ monochromatic radiation in a 2θ (diffraction angle) range from 20° to 120°. Additionally, to help distinguish the different chemical elements in any long-range order (LRO), and to quantify the average atomic displacement parameter (ADP), neutron powder diffraction patterns were measured by the high-resolution powder diffractometer for thermal neutrons with a wavelength of 1.1545 Å (HRPT diffractometer, Paul Scherrer Institute, PSI, Villigen, Switzerland). The neutron scattering lengths of Ta, Nb, Hf and Zr are 6.91, 7.05, 10.9 and 7.16 fm, respectively. So, ND provides the additional advantage of detecting any ordering between group 4 and group 5 elements (for example Hf and Ta), what would have been difficult by XRD due to the small atomic number contrast. Around 5 grams of powderized samples of HEA-2D and HEA-4D were put inside a cylindrical V container (5 mm diameter), and measureed for 4-5 hours each. To investigate into the local structural disorder and ADPs, single-crystal XRD datasets were collected in-house (Oxford Diffraction, Xcalibur) on crystals with less than 50 μm diameter using Mo $K_{\alpha}$ radiation. In order to determine the shape and intensities of the weak diffuse scattering arising from structural-disorder, single-crystal XRD datasets were collected using a PILATUS 2M detector at the Swiss-Norwegian Beamline (SNBL, beamline BM01A) at the European Synchrotron Research Facility (ESRF), Grenoble, France, with a wavelength of 0.6836 Å.

The microstructure and composition analyses were done with a scanning electron microscope (SEM, SU-70 Hitachi) in backscattered electron (BSE) mode equipped with an energy-dispersive X-ray spectroscope (EDX, X-MAX Oxford Instruments). The HEA samples (both as-cast and annealed) were cut and finely polished with a 60-nm $SiO_2$ suspension before investigating with SEM. For direct imaging of the lattice-fringes and local structural-disorder, a high-resolution transmission electron microscope (FEI Tecnai F30), equipped with a field emission gun operating at 300 keV was used. The HEA samples were first embedded inside a Cu tube, then cut into slices and polished with sand-papers to around 100 μm in thickness. Then, the discs were dimple-polished with 3-μm diamond suspension and finally milled by a precision Ar ion jet polishing system (Gatan PIPS II). Atom probe tomography (APT) was used (LEAP 4000X-HR system) to analyze the local chemical composition of any local disorder present. The probes for APT analysis were fabricated by focused ion-beam (FIB) milling by Ga ions in a FIB-SEM (Helios Nanolab 600i) system. The tips were conically tapered with a tip-radius of around 50 nm or less. The APT tips were then subjected to laser pulsing with 80-100 pJ of energy applied at 160 kHz inside the LEAP chamber during APT analysis. The charge state ratios of Ta, Nb, Hf and Zr were around 6.3, 1.2, 23 and 4.5, respectively for the APT runs. The local chemical composition analysis by APT eventually complemented the observations and understanding of the real local structure from HRTEM, XRD and ND. For the modeling of the real structure containing SRO/SRC, embedded-atom-method (EAM) potentials were used (for details see section 4).

The microhardness of the different HEAs was measured by a Vickers diamond indenter by applying 200 grams of weight. The measurements were done at 15 different positions scattered through the whole cross-section areas. Cylindrical specimens of 2.5 mm diameter and ~ 4.0 mm in length were cut out by electric-discharge machining for compression tests. A graphite lubricant was applied to the flat faces of the specimens, which were then compressed at an initial strain rate of $10^{-3}$ $s^{-1}$ until total fracture occurred. As a measure of ductility, the engineering strain at peak stress was considered. After reaching this strain level (peak strain), the materials generally start to develop internal cracks.

## 3. Results and discussion

### *3.1 Average structure and compositions:*

The as-cast and annealed TaNbHfZr alloys are single-phase with on average bcc structure as evident from the powder XRDs (Fig. 1). The experimental lattice parameter of the as-cast alloy is 3.431 ± 0.001 Å, which is close to the predicted value of 3.438 Å, according to Vegard's law [29] and 3.444 Å as predicted by Zen's law of average atomic volume [30]. As the hexagonal close packed (hcp) elements Hf and Zr are distributed in the bcc solid-solution phase, the expected bcc lattice parameters of these two elements were calculated from the atomic volumes of hcp phases at ambient-temperature [31]. The lattice parameters of the homogenized alloys (from HEA-6H to HEA-8D) were, however, around 3.39 Å, 1.2 % less than that of the as-cast alloy. For the HEA-4D, a distinct second minor hcp β phase forms, and this will be discussed later.

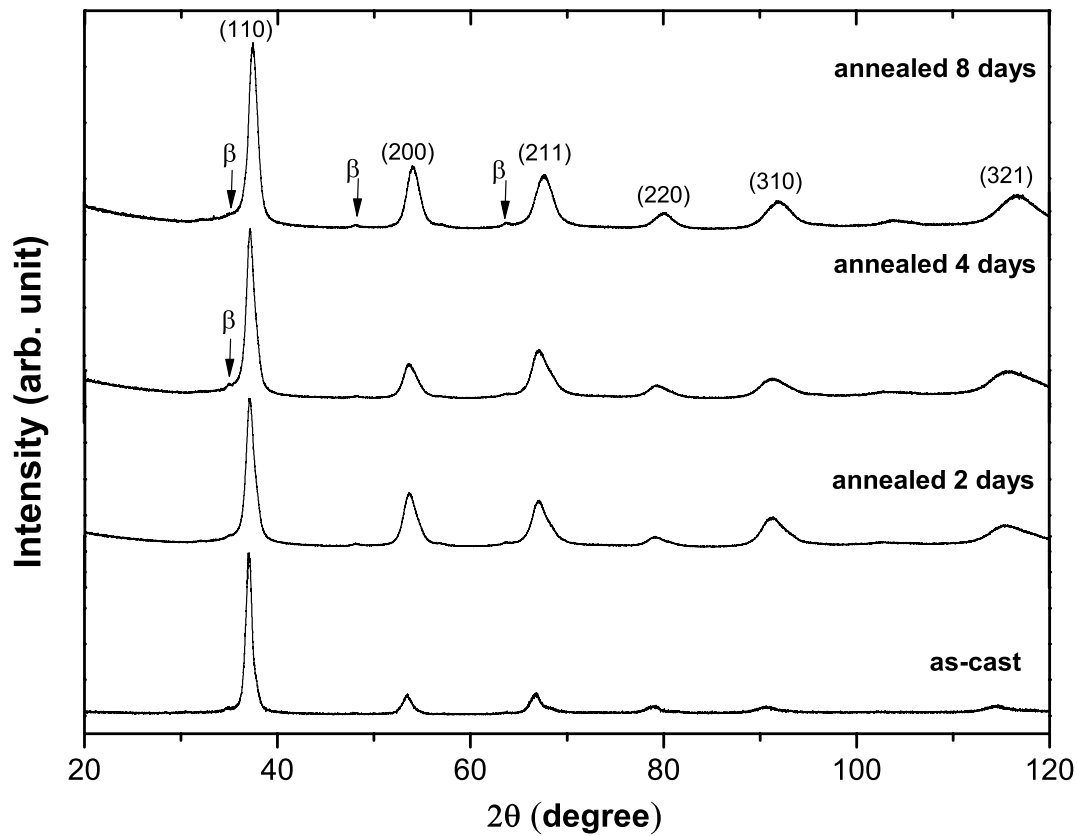

Fig. 1. Powder XRD patterns (Cu $K_{\alpha 1}$ radiation) of TaNbHfZr HEAs annealed for different times. Peaks from the bcc average structure are indexed on the top graph. Minor peaks from the hcp β phase are indicated by arrows.

Fig. 2 shows the SEM-BSE micrographs and the corresponding EDX line-scans across the HEAs annealed for different time periods. The as-cast (HEA-0D) alloy shows a dendritic microstructure (Fig 2a) with a maximum of 10 % (atomic proportion) elemental composition fluctuations for Ta and Zr, which are the elements with highest and lowest melting temperatures. This composition variation comes down to within 2 % for the HEA-6H (Figs. 2c and 2d). The EDX line-scan indicates that this 2 % composition variation gradually appears over a 50-100 μm distance. The microstructure (Fig. 2c) shows a homogeneous matrix, but also contains few traces of dissolving primary dendrite arms. For HEA-1D and HEA-4D (Figs. 2e and 2g) the composition only varies within 1% throughout the whole microstructure. The EDX line scans (Figs. 2f and 2h) running across the grain boundaries do not show any significant composition fluctuations.

*3.2 Local structural-disorder:*

The XRD peaks in Fig. 1 appear to be broader than some of the other refractory HEAs like MoNbTaW and MoNbTaWV [10, 16]. This was also found to be the case for powder ND on HEA-2D and HEA-4D. The ND patterns of the annealed HEAs do not reveal any extra features other than those of the XRD profiles. Fig. 3 shows the results of the Rietveld refinements of the ND pattern of HEA-2D. The refined value of the local internal-strain results to 1.062 ± 0.001 %, that of the coherent domain size to 10.094 ± 0.003 nm and the average ADP to 0.0185 ± 0.0003 Å$^2$. Similar values were also found from powder ND on HEA-4D. The refined average-structure ADPs of all the annealed samples are listed in Table 1. The neutron powder Rietveld refinement was done by *FullProf* [32] and the ADPs from the single crystal XRD datasets were refined by *SHELXL97* [33]. The ADPs of the annealed alloys, refined from the single-crystal XRDs fall in the range of 0.0214 - 0.0240 Å$^2$, whereas the ADPs from the powder ND experiments are in the range of 0.0185-0.0187 Å$^2$. These ADPs are clearly much larger than the expected thermal ADP, $U_t$, of 0.006 Å$^2$ calculated form the average of Debye-Waller factors of pure elements [34]. Similar results are also found in HEAs like MoNbTaW and ZrNbHf HEAs [16, 35]. The expected static component of the ADP, $U_s$, for this HEA amounts to 0.0188 Å$^2$, derived from the differences of lattice parameters of the pure elements as [16]:

$$U_s = \Sigma c_i (d_i - d_{al})^2$$

where, $c_i$ is the mole fraction of the element $i$, $d_i$ and $d_{al}$ are the lattice parameters of the pure elements and that of the HEA, respectively. The expected total ADP of the HEA, including thermal and static components is calculated to be $U_t + U_s$ = 0.006 + 0.0188 = 0.0248 Å$^2$, which is near to the experimental range of 0.0214 - 0.0240 Å$^2$ from the single-crystal XRDs. The ADPs refined from the powder NDs were little less than those of the single-crystal XRDs. This could be due to comparatively lesser amount of average displacements of the nuclei of the atoms than of the electron clouds around each nucleus. The average static atomic displacement, derived from the square root of the experimental static ADPs range between 0.124 - 0.134 Å.

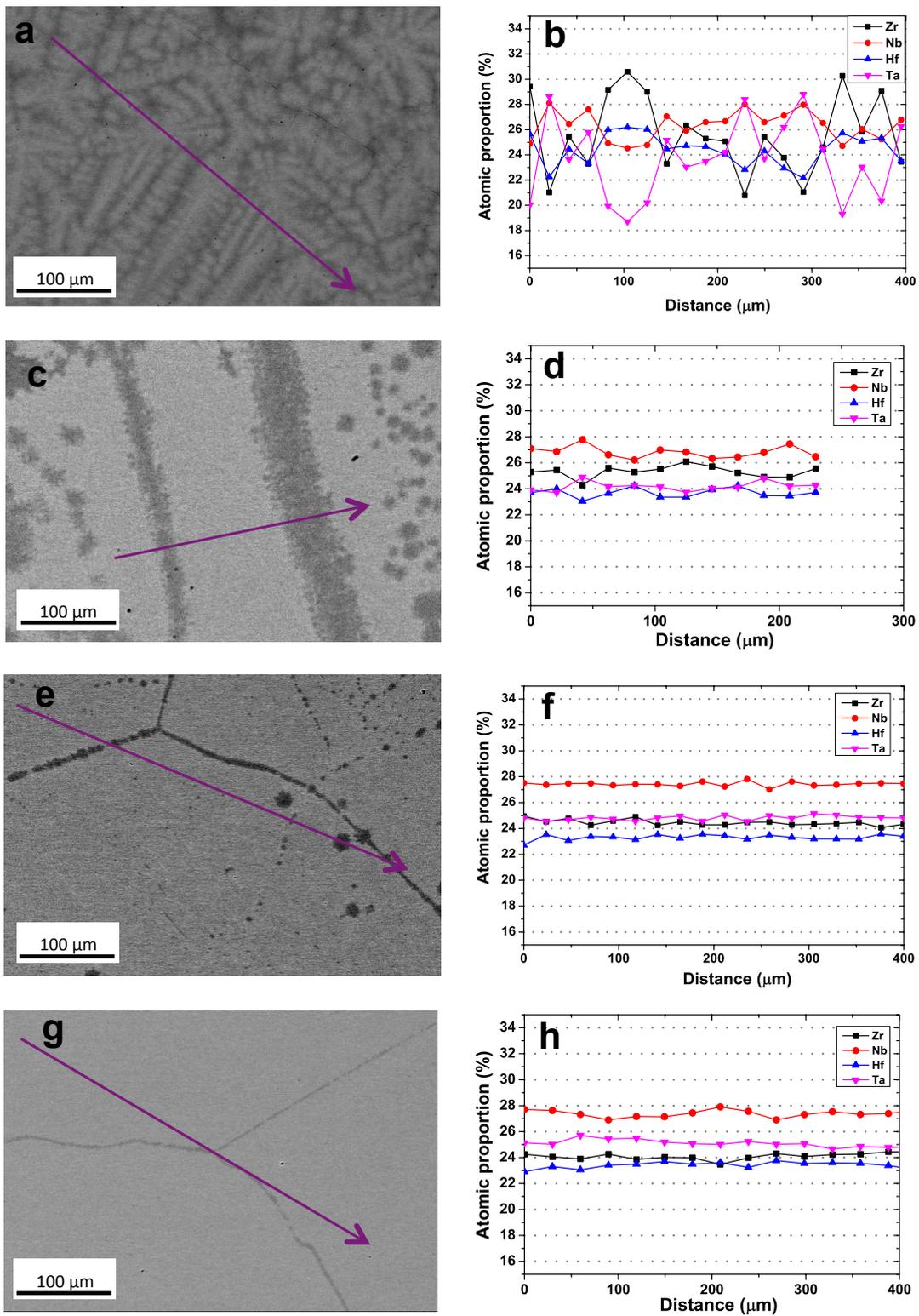

Fig. 2. BSE image of the cross-sectional microstructures of the HEA annealed for different times: (a) HEA-0D (as-cast), (c) HEA-6H, (e) HEA-1D and (g) HEA-4D. The arrows indicate the path of EDX line-scans. The EDX composition profiles are plotted in figures (b), (d), (f) and (h) for the different annealing conditions, respectively.

Table 1. Refined average-structure ADPs of HEA annealed for different times

| HEA sample | ADP from single-crystal XRD | ADP from powder ND |
| --- | --- | --- |
| HEA-6H | 0.0224 ± 0.0012 | na* |
| HEA-1D | 0.0227 ± 0.0016 | na |
| HEA-2D | 0.0214 ± 0.0012 | 0.0185 ± 0.0003 |
| HEA-4D | 0.0240 ± 0.0011 | 0.0187 ± 0.0004 |

* na: not available

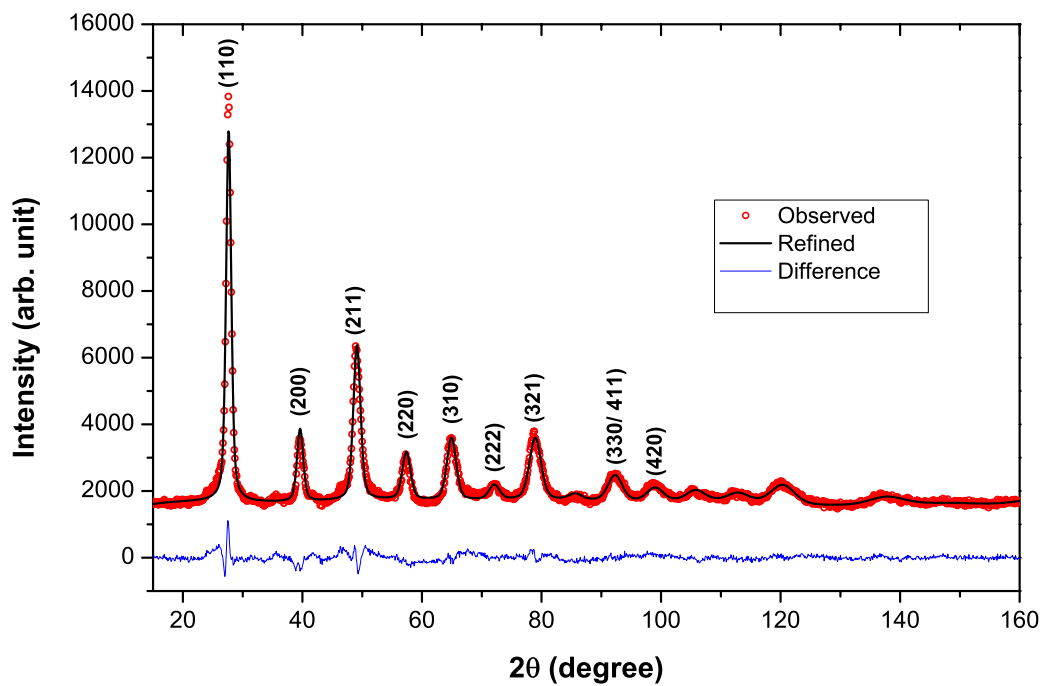

Fig. 3. Neutron powder diffraction pattern (data marks in red) of the HEA-2D with the peaks indexed for the bcc structure. The black line shows the Rietveld refinement and the differences in fitting are shown at the bottom in the blue line.

Since there were some weak streak-like diffuse scattering intensities observed near the Bragg reflections in the single-crystal XRD patterns, HRTEM investigations were done for the HEA-1D and HEA-4D alloys, specially to study the nature of local disorder. Fig. 4a and 4b depicts the bright field (BF) HRTEM images of HEA-1D taken along [100] zone axis. The micrographs (4a and 4b) show contrasts of adjacent lighter and darker regions appearing perpendicular to the main crystallographic axes. Also in the selected area electron diffraction (SAED) image (Fig. 4a, inset), the presence of streak-like diffuse scattering is observed in the direction of the [100] principal crystallographic axes. Asymmetric diffuse intensities around the Bragg reflections in SAED (indicated by the red arrows) and the contrasts in the BF images imply the presence of local disorder. Fig. 4b shows the lattice fringes of the local structures. The lattice fringes gradually bend in the areas of local image contrast (spreading perpendicular to the principal axes) due to tetragonal relaxations at the SRCs, which is discussed in detail in the following section on atom-probe tomography. The two boxed regions of Fig. 4b are enlarged in Figs. 4c and 4e, which contain the features of the local lattice relaxation and bending of the lattice fringes. Fast Fourier transforms (FFTs) of Figs. 4c and 4e are shown in Figs. 4d and 4f. The FFT images show streak-like features spreading out from the main Bragg spots towards the shorter reciprocal vectors similar to the SAED image in Fig. 4a. The diffuse streaks in the FFTs direct towards only one principal axis in each of the boxed regions of Fig. 4b, because mainly one set of a local SRC perpendicular to a particular [100] direction was included. Local tetragonal expansion relaxations and the diffuse intensity streaks corresponding to the SRCs indicate that the SRCs might have locally higher atomic volumes due to the clustering of some larger atoms.

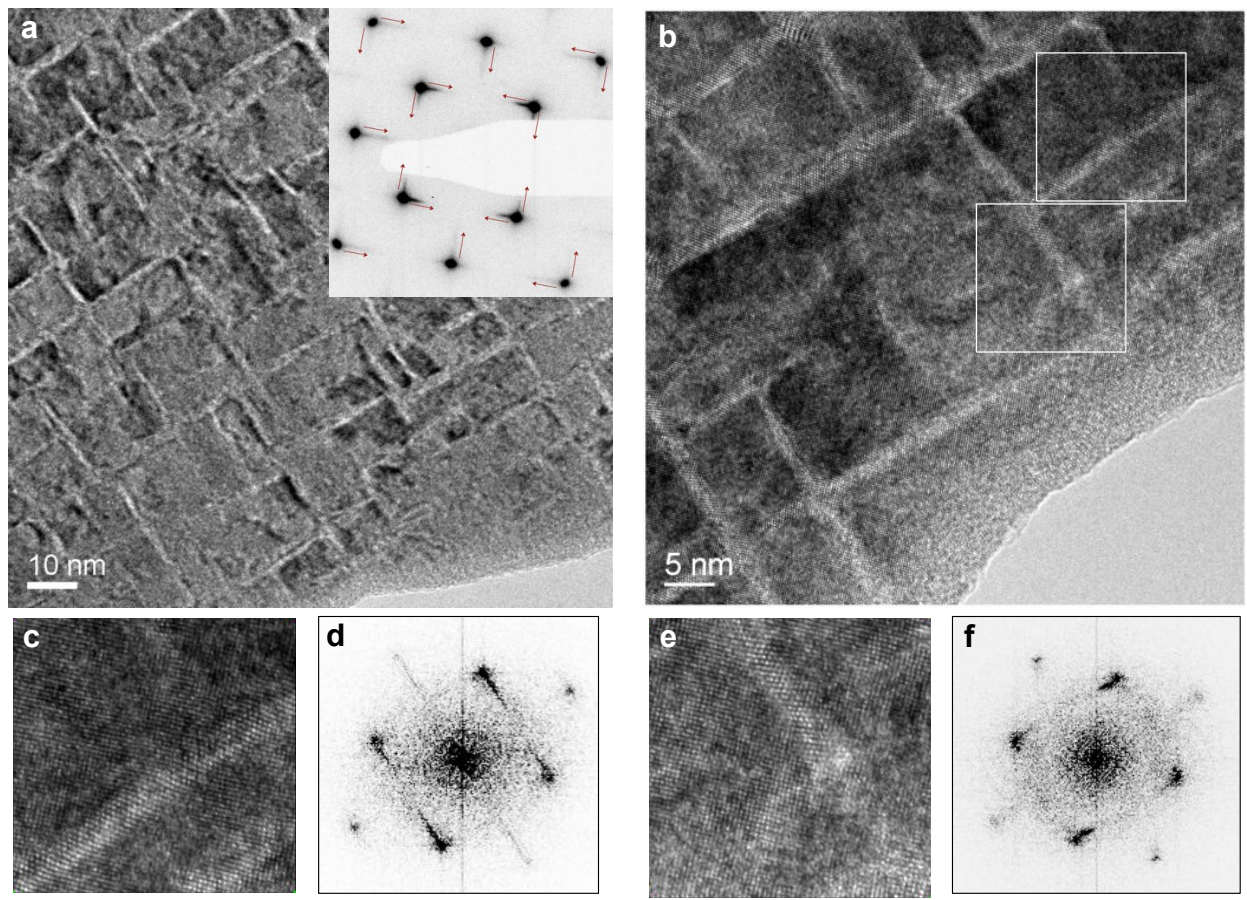

Fig. 4. HRTEM images of the HEA-1D: (a) TEM BF image along the [100] zone-axis. SAED pattern in the inset shows an asymmetric nature of streak-like diffuse scattering intensities indicated by small red arrows. (b) BF HRTEM image revealing lattice fringes and local lattice distortions at the SRCs. The two boxed regions are enlarged in (c) and (e) showing detailed view of local lattice relaxations/ distortions due to SRCs. (d) and (f) are the FFTs of (c) and (e), respectively, revealing the directions of asymmetric diffuse scatterings perpendicular to the two different set of SRCs.

The APT reconstructions of the elements were done for the samples HEA-0D, HEA-6H, HEA-1D and HEA-4D. Figs. 5a, 5c, 5e and 5g in the left column present the elemental maps of Zr, while Figs. 5b, 5d, 5f and 5h to the right show the elemental maps for Hf. The reconstructions from the as-cast alloy (HEA-0D in Figs 5a and 5b) show a homogeneous elemental distribution on the nano-scale. In contrast, the atom maps of the annealed alloys (HEA-6H, HEA-1D and HEA-4D) reveal some network-like clusterings of Zr and Hf. The HEA-6H alloy (Figs. 5c and 5d) had planer clusters of Zr, but the atomic clusters were rather isolated from each other. In the elemental map, there was some weak contrast of Hf atoms, but the in-depth concentration profiles (not shown here) did not reveal this weak clustering of Hf atoms. The clusters are separated by a distance of 6-8 nm from each other for the HEA-6H. However, for the HEA-1D the Zr clusters in the elemental map become more prominent and interconnected (Fig. 5e). The atom map of HEA-1D appears similar to that of the TEM micrographs in Fig. 4, where the interconnected perpendicular planar short-range clustering (SRC) is imaged in BF mode. Also, in the atom map for Hf (Fig. 5f) similar SRCs like that of Zr is observed, suggesting a trend of the larger Zr and Hf atoms to cluster together with longer annealing time. In the composition profile of HEA-1D across the SRCs, mainly Zr atoms cluster together while Hf atoms show only a weak tendency to co-cluster with Zr atoms. But, Hf co-cluster strongly with Zr in HEA-4D (Figs. 5g and 5h) as revealed in the in-depth concentration profiles. As the SRC planar clusters grow with time, they become more interconnected and make almost a 3d grid structure in HEA-4D (Fig. 5g). The spacing of the SRCs in HEA-1D and HEA-4D are in the range of 7-15 nm, which is again close to the spacings observed in HRTEM images (Figs. 4 and 7) and the calculated coherent domain sizes from the powder NDs. Thus, the prediction of clustering of bigger atoms and local SRC-disorder from the TEM investigation (Fig 4) is readily verified with the 3d APT reconstructions.

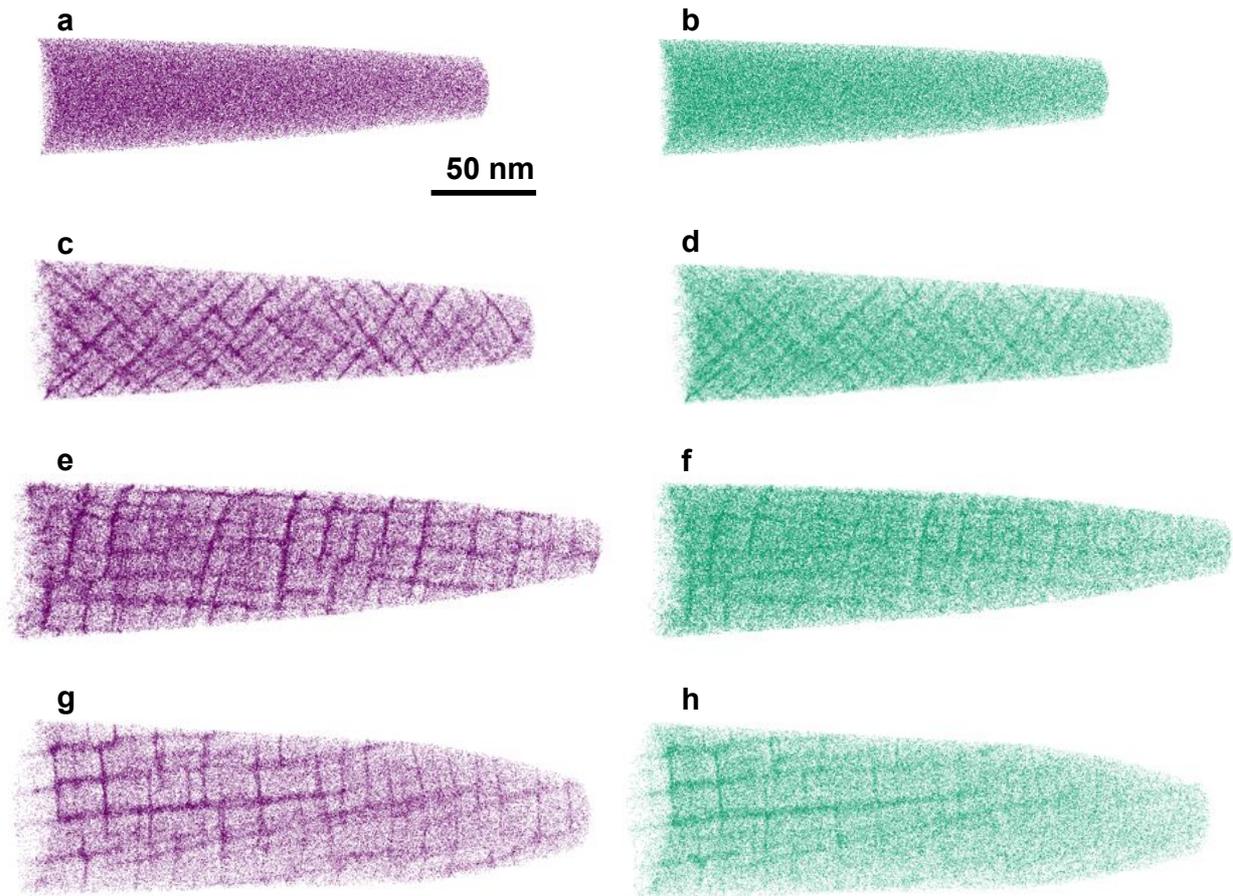

Fig. 5. Atom map reconstructions from APT analysis: (a), (c), (e) and (g) in the left column show the Zr atom maps for the alloys HEA-0D, HEA-6H, HEA-1D and HEA-4D, respectively. (b), (d), (f) and (h) to the right column show the Hf atom map for the same annealing conditions as described for the Zr atoms.

### *3.3 Modeling of the real structure and the local-disorder:*

The single-crystal XRD datasets collected by the PILATUS detector also reveal the presence of local-disorder based on the weak diffuse scattering. Figs. 6a and 6b show the reconstructed *hk0* reciprocal lattice layers of HEA-6H and HEA-4D. Similar to the SAED pattern (Fig 4a) and the FFT analysis of the HRTEM images (Figs 4d and 4f), the diffuse streaks adjacent to the Bragg reflections are asymmetric towards the origin. With higher reciprocal lattice vectors, the lengths of the streaks increase and reveal the presence of diffuse maxima within the streak. For the longer annealed sample HEA-4D, the diffuse intensities remain similar to that of the shorter annealed HEA-6H. But, some extra weak intensity spots (encircled spots) are observed for the HEA-4D, which correspond to the minor hcp β phase indicated earlier.

Fig. 6c shows the atomic model (relaxed structure) of the HEA containing a node-forming SRC with the maroon dots denoting Zr atoms and the white dots for the other elements. The atomic arrangement of the SRC is constructed from the APT local-composition analysis of the HEA-6H and HEA-1D material for this case. Details of the model are discussed later in section 4. Fig. 6d shows the experimental and simulated diffraction patterns of the annealed HEA model structure. The diffuse streaks get extended towards the shorter reciprocal vectors making a L-like shape in the *hk0* layer from both XRD and MD relaxations. The local tetragonal relaxation by the bigger Zr atoms causes the lattice layers of the MD model to bend at the SRCs (Fig. 6c). This explains the bending of lattice layers at the SRC as observed in the HRTEM images (Figs. 4c and 4e) causing the diffuse streak-like intensities extending from the Bragg spots. However, in the modeled structure the diffuse streaks do not extend to an extent as much as observed in the experiments. This can be expected taking into consideration that the EAM potentials can have limitations for large tetragonal deformations of around 30 % (expected from the *hk0* reconstructions of reciprocal space). A large degree of local tetragonal deformation at the SRCs is expected if the constancy of atomic volume for the elements is applied. In the MD relaxation process, atoms situated at the SRCs contribute significantly towards the minimization of internal energy of the structure, lowering it by 62 meV/atom compared to a statistically randomly distributed relaxed solid-solution. The procedure to determine the number of atoms at the SRCs contributing to the energy minimization is discussed in section 4. There was another interesting observation from the reconstructed *hk0* layers (Figs. 6a and 6b), the inherent mosaicity of the HEA crystals. The on average 1.6° mosaicity can be attributed to the degree of misalignments between the coherent-domains in a structure with SRCs frequently distributed throughout the whole volume. As the SRCs appear in the annealed HEAs, the matrix becomes depleted in the big Zr atoms decreasing its average lattice parameter compared to the as-cast alloy indicated earlier.

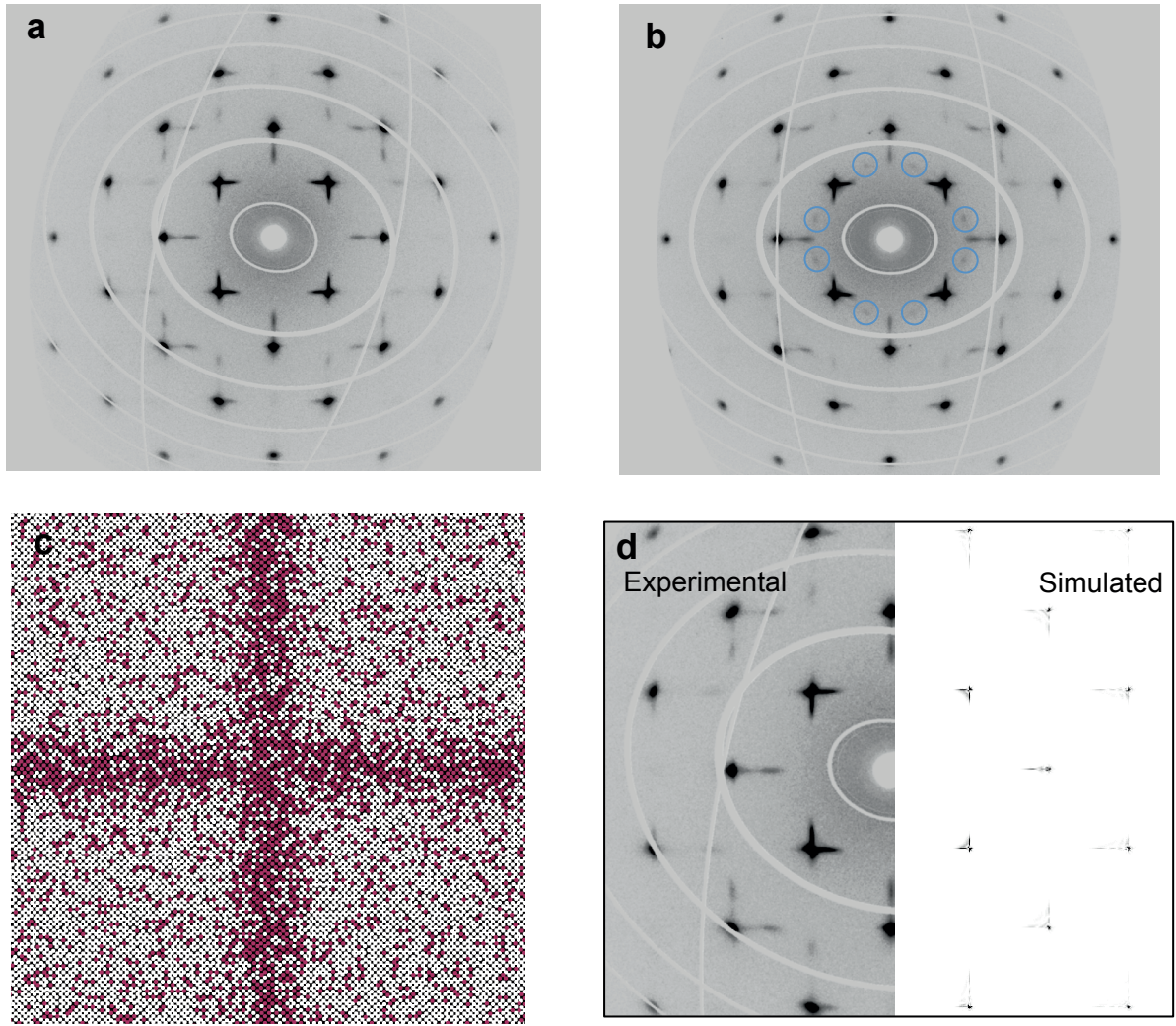

Fig. 6. Modeling of structural disorder by diffuse scattering: (a) and (b) are the reconstructed *hk0* reciprocal lattice layer from single-crystal XRD of the HEA-6H and the HEA-4D respectively, measured at the synchrotron beamline. In (b) the encircled spots are intensities from a minor hcp β phase. (c) Representative real structure model of the SRC where the maroon dots denote Zr and the other elements are shown in white. (d) Comparison of the diffuse scattering in *hk0* layer from the single-crystal XRD (left) and (right) from the MD relaxed atomic model in (c).

*3.4 Structural evolution with annealing time:*

As the SRCs become more interconnected with longer annealing times, and a minor hcp β phase appears after 4 days, it is anticipated that these would influence the structural and mechanical properties of the HEA. Some more HRTEM investigations have been done for the HEA-4D sample in a similar fashion to that of HEA-1D (Fig. 4). Figs. 7a and 7b show the high-resolution BF images taken along the [100] crystallographic zone-axis with the SAED pattern given in the inset of Fig. 7a. The image contrast is similar to the BF HRTEM images of HEA-1D showing cuboid-like domains and also to that of the APT reconstructions (Figs. 5e and 5g). The main difference of the local structure of HEA-4D from HEA-1D is that the SRCs become more interconnected with each other, and in the SRCs Hf atoms cluster strongly together with Zr atoms. In HEA-4D, the SRCs make a 3d network meeting each other at some cross-like clustering nodes and thus creates grid-like repeating domains of cuboids all throughout the structure. Diffraction spots from the minor hcp β phase are observed in the SAED for HEA-4D (Fig. 7c) similar to the *hk0* reconstructed layer from the single-crystal XRD (Fig. 6b), given that a longer exposure time (5-10 seconds) for the SAED pattern is provided. These extra weak diffraction spots of the β phase are indexed with orange color lines along with the strong Bragg intensities of the average bcc matrix (Fig. 7c). The β phase apparently has multiple twin components and two of them are shown, one with solid and another with dashed orange lines. One of the stronger diffraction spots of the β phase (encircled spot in Fig. 7c) was taken for a dark field (DF) imaging as shown in Fig. 7d. The DF image reveals that the selected Bragg intensity is originating from the nodes of the SRCs, indicating that the hcp β phase forms at the high concentration regions of the hcp elements Hf and Zr. However, there are some traces of this β phase also in SRC regions away from the clustering-nodes. It appears that with longer than 1 day of annealing, the SRCs grow and form clustering nodes (as in HEA-4D), where it reaches the critical size for the nucleation of the hcp β phase.

The HEA-8D sample has even a higher degree of this β phase precipitation as suggested by the increased peak intensity in powder XRD (Fig. 1). HEA-8D shows decomposition of the average structure into two different bcc phases with close lattice parameters, as observed in single-crystal XRDs (not shown here). This could be due to the tendency that the cuboid regions of SRCs become more enriched in Ta and Nb, forming a connected sub-micron scale long-range average structure in HEA-8D. Detailed investigations into the structural-decomposition of HEA-8D are out of the scope of this work.

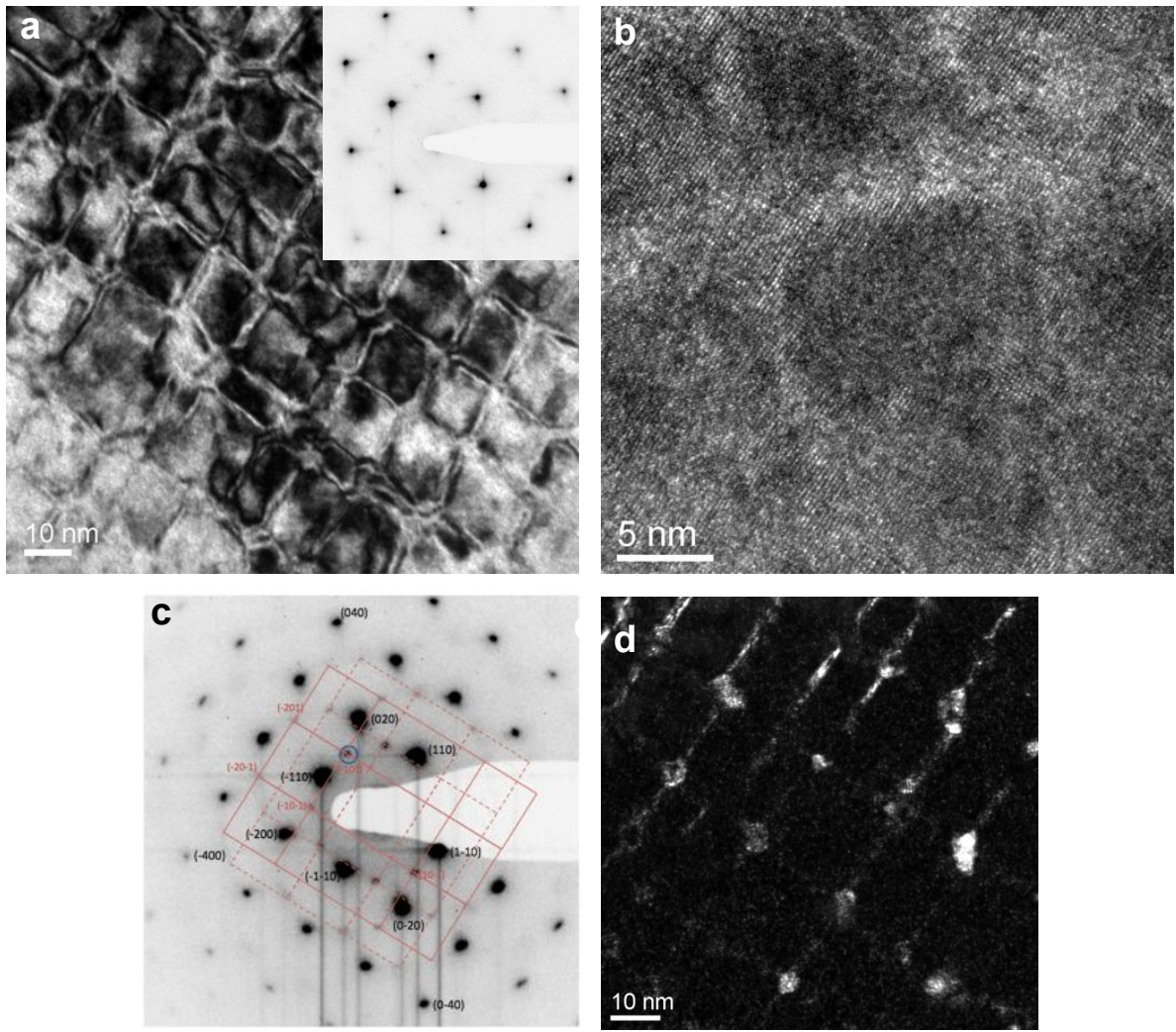

Fig. 7. HRTEM images of the HEA-4D: (a) BF image along [100] zone-axis with SAED pattern in the inset. (b) BF HRTEM image revealing lattice fringes at the SRCs and the clustering nodes. (c) SAED pattern with a longer exposure where the Bragg reflections from the minor hcp β phase are indexed with the orange color lines and reflections from the bcc matrix phase are shown in black. (d) DF image taken from the β phase reflection encircled in the SAED pattern in (c). DF image contrast reveals the presence of the β phase at the larger connecting nodes of the SRCs.

### *3.5 Mechanical properties evolution with annealing time:*

Fig. 8 shows the evolution of hardness, compressive yield strength and ductility (peak strain) from the as-cast to the 8 days annealed HEAs. The hardness and the compressive yield strength of the HEA in the as-cast condition are 3575 MPa and 1315 MPa, which are 2.4 and 4.9 times, respectively, of what is expected from the rule of mixture. The hardness increases to a level of 5598 MPa for the HEA-1D. Similar to the increase of hardness, the yield strengths of the materials also show an increase from 1315 MPa to 2310 MPa for HEA-1D (Fig. 8b), accompanied by a rapid drop in peak strain from

21.6 % down to 0.35 %. Similar to this TaNbHfZr HEA, a 2-3 times increase in hardness of the as-cast HEAs are also observed in MoNbTaW and MoNbTaWV [9, 16]. High hardness and strength of the HEAs even in the as-cast conditions with no detectable SRO/ SRCs present, are mainly attributed to the local lattice distortion and solid-solution strengthening effects [9, 16, 35]. Apart from the solid solution strengthening in the as-cast condition, with the increase in annealing time and the density of SRCs, the hardness and yield strength of the TaNbHfZr HEA increases by 57 % and 76 %, respectively, for HEA-1D. SRCs and GP zones formed due to annealing treatments have been found to additionally increase the yield strength and reduce the ductility in many substitutional Al [22] and Ni based alloys [28]. During early stages of annealing in HEA-6H, SRCs appear in larges numbers increasing the strength by 424 MPa and reducing the peak strain to 2.25 %. As the SRCs become increasingly interconnected from the as-cast to the 1 day annealed state, any dislocation moving through the material will be subjected to overcome the local energy minima caused by the SRCs. This causes the yield strength to reach a maximum for the HEA-1D. The increase of yield strength due to SRCs can be described for interstitial [26] and substitutional alloys [36] by the following relations:

$$\sigma_{SRC} = M \left( \frac{E_{Random} - E_{SRC}}{b^3} \right)$$

where, $\sigma_{SRC}$ is the increase in yield strength due to SRC, $M$ is Taylor factor ≈ 2.75 for bcc structures [37], $E_{Random}$ and $E_{SRC}$ are the energies (per atom) of the relaxed structure of the random solid solution matrix and the SRCs, $b$ is the Burgers vector equal to 2.936 Å. The $E_{Random}$ - $E_{SRC}$ value is obtained from the structure modeling by MD relaxations described in the section 4. The calculated value of $\sigma_{SRC}$ from the given model is found to be 1079 MPa, which is 8.44 % higher than the experimental value of (2310 – 1315) MPa = 995 MPa. This difference can arise from the errors of calculating $E_{SRC}$ and due to the relatively large scatter in the measured values of yield strength of the brittle HEA-1D material.

From a peak of hardness and yield strength for the HEA-1D, the values drop by around 6 % and 13 %, respectively for the HEA-8D (Fig 8). The peak strain also increases from a near-zero for the HEA-1D upto 3.5 % for the HEA-8D. This can be explained by the fact that around 4 days of annealing, the SRCs all become interconnected and the clustering nodes grow to accommodate the hcp Zr/Hf rich β phase. This in turn would reduce the internal strain in the matrix phase and ease the movement of dislocations to allow for some plastic deformations and gradually improve the ductility beyond 2-4 days of annealing. However, as the HEA-8D material experiences the decomposition of the bcc single-phase matrix into two bcc phases, there could be a further lowering of the internal strain, which would improve ductility. This mechanism is also supported from the experimental data, but since this work mainly deals with the single-phase HEAs, detailed investigations of the HEA-8D is out of the scope of this work.

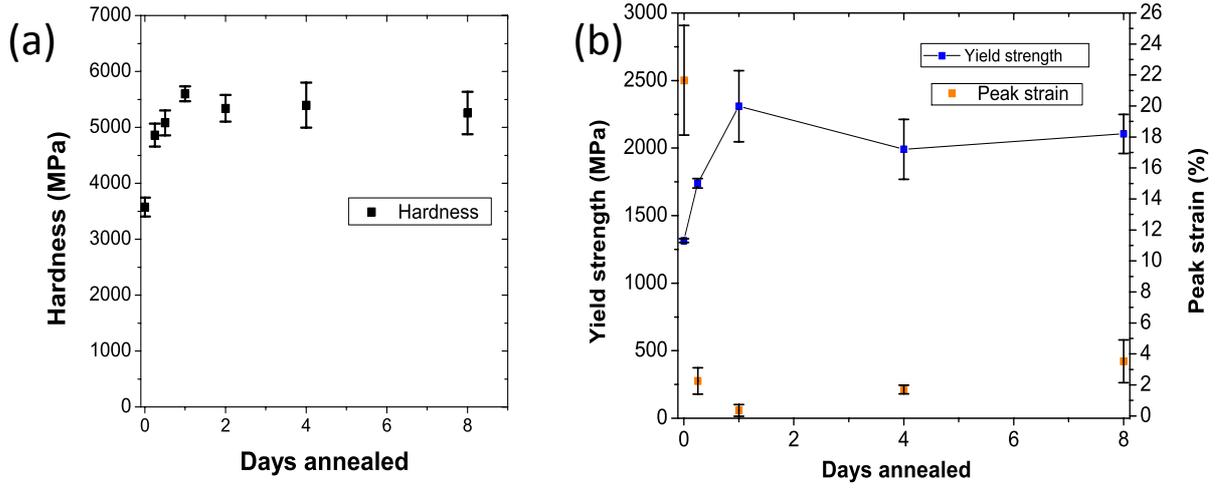

Fig. 8. Mechanical property evolution with annealing time: (a) Vickers microhardness for different annealing times. (b) Yield strength and ductility (engineering peak-strain).

## 4. Structure modeling

An atomic model of the HEA was constructed depicting the local structural disorder in a simulation volume of 80×80×20 supercells of the bcc average structure containing a total of 256000 atoms. EAM potentials [38, 39], suitable for bcc metals and alloys [40], were used for the MD relaxations. In the framework of EAM potentials, the total energy of the system is given by the basic equations:

$$E_t = \sum_i F_i(\rho_i) + \frac{1}{2} \sum_{i,j\ (i \neq j)} \varphi_{ij}(r_{ij}),$$

$$\rho_i = \sum_{i(i \neq j)} f_j(r_{ij}),$$

where $E_t$ is the total internal energy of the system, $F_i(\rho_i)$ is the energy to embed atom $i$ in an electron density $\rho_i$, $\rho_i$ is the electron density at atom $i$ due to all other atoms given by $f_j(r_{ij})$, $r_{ij}$ is the distance between atom $i$ and $j$, and $\varphi_{ij}(r_{ij})$ is a two-body interaction potential between atoms $i$ and $j$. The atomic interaction potentials were created based on physical parameters of the elements such as atomic radius, cohesive energy, unrelaxed vacancy formation energy and second order elastic stiffness constants. These physical input parameters were suitably taken for a bcc structure of the elements and presented in Table 2. MD relaxation was done with the simulation program *LAMMPS* [41] by using the conjugate gradient method. The internal energy of the system was minimized to a tolerance level of $10^{-14}$ for the atomic relaxations. Since the average structure of the HEA remained a bcc solid solution in all the annealing conditions, the stiffness constants of the elements were taken from bcc phases as well.

Table 2. Input physical parameters of elements for the construction of the EAM potentials

| Element | Atomic radius (Å) | $E_c$ (eV) | $E_f$ (eV) | $C_{11}$ (eV/ Å$^3$) | $C_{12}$ (eV/ Å$^3$) | $C_{44}$ (eV/ Å$^3$) |
|---|---|---|---|---|---|---|
| Ta | 1.430[a] | 9.71[b] | 2.95[a] | 1.648[e] | 0.986[e] | 0.516[e] |
| Nb | 1.429[a] | 8.21[b] | 2.75[a] | 1.529[e] | 0.824[e] | 0.177[e] |
| Hf | 1.572[c] | 7.56[b] | 2.39[d] | 0.818[f] | 0.643[f] | 0.281[f] |
| Zr | 1.608[c] | 6.31[b] | 2.30[d] | 0.649[g] | 0.580[g] | 0.237[g] |

$E_c$ and $E_f$ are the cohesive and unrelaxed vacancy formation energies.

References: a [38], b [42], c [31], d [43], e [44], f [45], g [46].

The in-depth APT composition profile across the SRCs of HEA-6H and HEA-1D, generally has a Gaussian-like distribution of Zr with a peak composition between 60-75 % and a FWHM of around 2.0-2.6 nm. For the current model, Hf atoms are randomly distributed throughout the local structure and the SRCs are depleted in Ta and Nb concentration as found in the HEAs. The atomic model has a Gaussian type Zr concentration with a peak composition of 67.5 % and a FWHM of 2.5 nm. The model atomic structure was then relaxed with the developed EAM potential and a diffraction pattern of the relaxed structure was calculated using the *DISCUS* package [47].

While calculating $E_{SRC}$ for the energy minimization by the SRC, it is assumed that the atoms located within full width at tenth maximum (FWTM) of the Gaussian (around 97 % of area under the Gaussian curve) composition distribution significantly affect the energy minimization process. In the literature some of the $E_{SRC}$ values have been calculated by ab-initio methods, where the SRCs are contained within a volume of a few hundred atoms [26]. But, for more diffuse and extended SRCs like in this HEA, the calculation of $E_{SRC}$ could be nontrivial. However, it has been shown by the ab-initio based studies of SRO and SRCs in binary alloys that the significant minimization in energy occurs after the local structure relaxation [48]. For this HEA, the energy difference between the unrelaxed structures of a random solid-solution and that containing the SRC is negligible. Whereas, the relaxed structures show a marked 62 meV/atom of internal energy difference between the $E_{Random}$ and the $E_{SRC}$.

# 5. Summary and conclusions

The TaNbHfZr HEA shows a bcc average structure in the as-cast and the long-term annealed condition. The average structure reflects the high degree of local lattice distortion by the large static ADP based on the single-crystal XRD and the powder ND experiments. With longer annealing time at 1800°C, an increasing degree of SRC perpendicular to the principal axes is observed. The APT analysis reveals that the local composition of the SRC is high in Zr and depleted in Ta and Nb. But, after 4 days of annealing Hf and Zr cluster together in the SRCs. The SRCs cause local tetragonal lattice relaxation as evidenced from the direct HRTEM imaging and the diffuse scattering intensities in SAED and single-crystal XRD. The asymmetric streak-like nature of the diffuse scattering intensities could be partially reproduced by the MD relaxations of a model structure containing SRCs. With longer annealing times, the SRCs become more interconnected forming clustering nodes and grid-like repeating domains. The as-cast yield strength (1315 MPa) of the HEA is 4.9 times that expected from the expected value due to a solid-solution-strengthening-like mechanism. But, the peak of yield strength (2310 MPa) and hardness accompanied by a minimum of ductility is observed after 1 day of annealing. The increase in yield strength could be adequately explained with the local energy minimization by SRCs as calculated by the MD relaxation of a model structure. After 4 days of annealing, nucleation of a hcp β phase is evidenced at the larger nodes of the SRCs as observed from the XRD and the HRTEM studies. This β phase precipitation lowers the internal stress of the matrix, causing the yield strength to drop and ductility to improve again beyond 1 day of annealing. Our study reveals the details of the local disorder present, their mechanism of evolution with long annealing time and its effect on the mechanical properties at ambient-temperature.

**Acknowledgements**

The authors would like to thank Dr. S. Gerstl, Dr. F. Gramm and Dr. R. Schäublin (SCOPEM, ETH Zurich) for their technical help in APT and HRTEM; Dr. D. Chernyshov (SNBL, ESRF Grenoble) for the support in the single-crystal XRD at the synchrotron beamline. The authors also gratefully acknowledge the financial support of Swiss National Science Foundation under the grant 200020_144430.

# References

[1] J.W. Yeh, S.K. Chen, S.J. Lin, J.Y. Gan, T.S. Chin, T.T. Shun, C.H. Tsau, S.Y. Chang, Nanostructured high-entropy alloys with multiple principal elements: Novel alloy design concepts and outcomes, Advanced Engineering Materials 6 (2004) 299-303.

[2] X. Yang, Y. Zhang, Prediction of high-entropy stabilized solid-solution in multi-component alloys, Materials Chemistry and Physics 132 (2012) 233-238.

[3] M.S. Lucas, G.B. Wilks, L. Mauger, J.A. Munoz, O.N. Senkov, Absence of long-range chemical ordering in equimolar FeCoCrNi, Applied Physics Letters 100 (2012) 251907.

[4] Y.J. Zhou, Y. Zhang, F.J. Wang, G. L. Chen, Phase transformation induced by lattice distortion in multiprincipal component $CoCrFeNiCu_xAl_{1-x}$ solid-solution alloys, Applied Physics Letters 92 (2008) 241917.

[5] K.Y. Tsai, M.H. Tsai, J.W. Yeh, Sluggish diffusion in Co–Cr–Fe–Mn–Ni high-entropy alloys, Acta Materialia 61 (2013) 4887-4897.

[6] Y. Zhang, T.T. Zuo, Z. Tang, M.C. Gao, K.A. Dahmen, P.K. Liaw, Z.P. Lu, Microstructures and properties of high-entropy alloys, Progress in Materials Science 61 (2014) 1-93.

[7] S.Y. Chang, C.E. Li, Y.C. Huang, H.F. Hsu, J.W. Yeh, S.J. Lin, Structural and Thermodynamic Factors of Suppressed Interdiffusion Kinetics in Multi-component High-entropy Materials, Scientific Reports 4 (2014) 4162.

[8] B. Gludovatz, A. Hohenwarter, D. Catoor, E.H. Chang, E.P. George, R.O. Ritchie, A fracture-resistant high-entropy alloy for cryogenic applications, Science 345 (2014) 1153-1158.

[9] O.N. Senkov, G.B. Wilks, D.B. Miracle, C.P. Chuang, P.K. Liaw, Refractory high-entropy alloys, Intermetallics 18 (2010) 1758-1765.

[10] O.N. Senkov, G.B. Wilks, J.M. Scott, D.B. Miracle, Mechanical properties of $Nb_{25}Mo_{25}Ta_{25}W_{25}$ and $V_{20}Nb_{20}Mo_{20}Ta_{20}W_{20}$ refractory high entropy alloys, Intermetallics 19 (2011) 698-706.

[11] O.N. Senkov, J.M. Scott, S.V. Senkova, D.B. Miracle, C.F. Woodward, Microstructure and room temperature properties of a high-entropy TaNbHfZrTi alloy, Journal of Alloys and Compounds 509 (2011) 6043-6048.

[12] O.N. Senkov, J.M. Scott, S.V. Senkova, F. Meisenkothen, D.B. Miracle, C.F. Woodward, Microstructure and elevated temperature properties of a refractory TaNbHfZrTi alloy, Journal of Materials Science 47 (2012) 4062-4074.

[13] O.N. Senkov, S.V. Senkova, C. Woodward, D.B. Miracle, Low-density, refractory multi-principal element alloys of the Cr–Nb–Ti–V–Zr system: Microstructure and phase analysis, Acta Materialia 61 (2013) 1545-1557.

[14] D. Kaufmann, R. Mönig, C.A. Volkert, O. Kraft, Size dependent mechanical behaviour of tantalum, International Journal of Plasticity 27 (2011) 470-478.

[15] G.Z. Voyiadjis, A.H. Almasri, T. Park, Experimental nanoindentation of BCC metals, Mechanics Research Communications 37 (2010) 307-314.

[16] Y. Zou, S. Maiti, W. Steurer, R. Spolenak, Size-dependent plasticity in an $Nb_{25}Mo_{25}Ta_{25}W_{25}$ refractory high-entropy alloy, Acta Materialia 65 (2014) 85-97.

[17] M. Widom, W.P. Huhn, S. Maiti, W. Steurer, Hybrid Monte Carlo/Molecular Dynamics Simulation of a Refractory Metal High Entropy Alloy, Metallurgical and Materials Transactions A 45 (2014) 196-200.

[18] M. Hillert, A solid-solution model for inhomogeneous systems, Acta Metallurgica 9 (1961) 525-535.

[19] E. Metcalfe, J.A. Leake, An X-ray diffuse scattering study of short-range order in CuAu, Acta Metallurgica 23 (1975) 1135-1143.

[20] J. Vrijen, S. Radelaar, Clustering in Cu-Ni alloys: A diffuse neutron-scattering study, Physical Review B 17 (1978) 409-421.

[21] K. Osaka, T. Takama, X-ray study of the short-range order structure in Cu–24.3 at.% Mn alloy, Acta Materialia 50 (2002) 1289–1296.

[22] J.B. Cohen, The Internal Structure of Guinier-Preston Zones in Alloys, Solid State Physics 39 (1986) 131-206.

[23] S.M. Dubiel, J. Zukrowski, Phase-decomposition-related short-range ordering in an Fe-Cr alloy, Acta Materialia 61 (2013) 6207-6212.

[24] B.D. Butler, J.B. Cohen, Atomic displacements caused by carbon interstitials in austenite, Acta Metallurgica et Materialia 41 (1993) 41-48.

[25] V.G. Gavriljuk, B.D. Shanina, H. Berns, On the correlation between electron structure and short range atomic order in iron-based alloys, Acta Materialia 48 (2000) 3879-3893.

[26] J.H. Kang, T. Ingendahl, J. von Appen, R. Dronskowski, W. Bleck, Impact of short-range ordering on yield strength of high manganese austenitic steels, Materials Science & Engineering A 614 (2014) 122–128.

[27] N.R. Dudova, V.A. Valitov, R.O. Kaibyshev, Short-Range Order and Mechanical Properties of Nichrome, Technical Physics 54 (2009) 77-79.

[28] R.L. Klueh, J.F. King, Short-Range Order Effects on the Tensile Behavior of a Nickel-Base Alloy, Metallurgical Transactions A 10 (1979) 1543-1548.

[29] A.R. Denton, N.W. Ashcroft, Vegard's law, Physical Review A 43 (1991) 3161-3164.

[30] E.A. Zen, Validity of Vegard's Law, American Mineralogist 41 (1956) 523-524.

[31] L.C. Ming, M.H. Manghnani, K.W. Katahara, Investigation of α→ω transformation in the Zr-Hf system to 42 GPa, Journal of Applied Physics 52 (1981) 1332-1335.

[32] J. Rodriguez-Carvajal, Recent advances in magnetic structure determinationby neutron powder diffraction, Physica B 192 (1993) 55-69.

[33] G.M. Sheldrick, A short history of SHELX, Acta Crystallographica Section A 64 (2008) 112-122.

[34] L.M. Peng, G. Ren, S.L. Dudarev, M.J. Whelan, Debye-Waller Factors and Absorptive Scattering Factors of Elemental Crystals, Acta Crystallographica Section A 52 (1996) 456-470.

[35] W. Guo, W. Dmowski, J.Y. Noh, P. Rack, P.K. Liaw, T. Egami, Local Atomic Structure of a High-Entropy Alloy: An X-Ray and Neutron Scattering Study, Metallurgical and Materials Transactions A 44 (2013) 1994-1997.

[36] M.J. Starink, L.F. Cao, P.A. Rometsch, A model for the thermodynamics of and strengthening due to co-clusters in Al–Mg–Si-based alloys, Acta Materialia 60 (2012) 4194–4207.

[37] J.M. Rosenberg, H.R. Piehler, Calculation of the Taylor Factor and Lattice Rotations for Bcc Metals Deforming by Pencil Glide, Metallurgical Transactions 2 (1971) 257-259.

[38] R.A. Johnson, D.J. Oh, Analytic embedded atom method model for bcc metals, Journal of Materials Research 4 (1989) 1195-1201.

[39] X.W. Zhou, H.N.G. Wadley, R.A. Johnson, D.J. Larson, N. Tabat, A. Cerezo, A.K. Petford-Long, G.D.W. Smith, P.H. Clifton, R.L. Martens, T.F. Kelly, Atomic scale structure of sputtered metal multilayers, Acta Materialia 49 (2001) 4005-4015.

[40] Z. Bangwei, O. Yifang, Theoretical calculation of thermodynamic data for bcc binary alloys with the embedded-atom method, Physical Review B 48 (1993) 3022-3029.

[41] S. Plimpton, Fast Parallel Algorithms for Short-Range Molecular Dynamics, Journal of Computational Physics 117 (1995) 1-19.

[42] G. Kaptay, G. Csicsovszki, M.S. Yaghmaee, An Absolute Scale for the Cohesion Energy of Pure Metals, Materials Science Forum 414-415 (2003) 235-240.

[43] O.L. Bacq, F. Willaime, A. Pasturel, Unrelaxed vacancy formation energies in group-IV elements calculated by the full-potential linear muffin-tin orbital method: Invariance with crystal structure, Physical Review B 59 (1999) 8508-8515.

[44] Y. Ouyang, B. Zhang, Analytic embedded-atom potentials for bcc metals: application to calculating the thermodynamic data of bcc alloys, Physics Letters A 192 (1994) 79-86.

[45] J. Trampenau, A. Heiming, W. Petry, M. Alba, C. Herzig, W. Miekeley, H.R. Schober, Phonon dispersion of the bcc phase of group-IV metals. III. bcc hafnium, Physical Review B 43 (1991) 10963-10969.

[46] A. Heiming, W. Petry, J. Trampenau, M. Alba, C. Herzig, H.R. Schober, G. Vogl, Phonon dispersion of the bcc phase of group-IV metals. II. bcc zirconium, a model case of dynamical precursors of martensitic transitions, Physical Review B 43 (1991) 10948-10962.

[47] T. Proffen, R.B. Neder, DISCUS, a Program for Diffuse Scattering and Defect Structure Simulations, Journal of Applied Crystallography 30 (1997) 171-175.

[48] C. Wolverton, V. Ozolins, A. Zunger, First-principles theory of short-range order in size-mismatched metal alloys: Cu-Au, Cu-Ag, and Ni-Au, Physical Review B 57 (1998) 4332-4348.